\documentclass{iau}

\usepackage{amsmath}
\usepackage{graphicx}
\usepackage{multirow}

\begin{document}

\lefttitle{A. Sterling et al.}
\righttitle{Solar Coronal Jets, and Implications for CME-Producing Eruptions}

\jnlPage{1}{7}
\jnlDoiYr{2024}
\doival{10.1017/xxxxx}
\volno{388}
\pubYr{2024}
\journaltitle{Solar and Stellar Coronal Mass Ejections}

\aopheadtitle{Proceedings of the IAU Symposium}
\editors{N. Gopalswamy,  O. Malandraki, A. Vidotto \&  W. Manchester, eds.}

\title{On the Onset Mechanism for Solar Coronal Jets, and Implications for the Onset Mechanism for CME-Producing Eruptions}

\author{Alphonse C. Sterling$^1$, Ronald L. Moore$^{2,1}$ \& Navdeep K. Panesar$^{3,4}$}
\affiliation{(1) NASA Marshall Space Flight Center, Huntsville, AL, USA; (2) The University of Alabama in Huntsville, Huntsville, AL, USA.; (3) Bay Area Environmental Research Institute, Moffett Field, CA, USA; (4) Lockheed Martin Solar and Astrophysics Laboratory, Palo Alto, CA, USA}

\begin{abstract}
Large-scale solar eruptions often include ejection of a filament, a solar flare, and expulsion of a coronal mass ejection (CME). Unravelling the magnetic processes that build up the free energy for these eruptions and trigger that energy's release in the eruption is a continuing challenge in solar physics. Such large-scale eruptions are comparatively infrequent, with the moderate level ones (say, GOES M-class events) occurring perhaps once every few days on average during active-activity times, and much less frequently during quieter times. In contrast, solar coronal jets, which are long ($\sim$50,000 km), narrow (less than about 10,000 km), transient ($\sim$10---20 min) plasma spires with bright bases and that are seen in soft X-rays and EUV, occur much more frequently, likely several hundred times per day independent of large-scale solar activity level. Recent studies indicate that coronal jets are small-scale versions of large-scale eruptions, often produced by eruption of a small-scale ``miniflament," that results in a ``miniflare" analogous to a larger typical solar flare, and that sometimes produces a CME analogue (a ``narrow CME" or ``white-light jet"). Under the assumption that jets are small-scale eruptions, their higher occurrence frequency and faster build-up evolution reveals perhaps fundamental aspects of all eruptions that are not as easy to discern in the more-complex magnetic environment and the slower build up to the larger eruptions. Therefore, the study of coronal jets can provide insights into the onset mechanism of CME-producing large-scale eruptions.
\end{abstract}


\maketitle

\section{Overview of Solar Corona Jets}

Solar coronal jets are long, narrow geyser-like ejections usually observed in soft X-ray (SXR) or EUV coronal images (Shibata et al.\ 1992, Raouafi et~al.~2016, Shen~2021, Schmieder~2022, Sterling et al.\ 2023). They reach $\sim$50,000 km, have widths $\sim$8000 km, and lifetimes $\sim$10 min 
(Savcheva et~al.~2007). (Fig. 1.)

\begin{figure} 
\hspace*{3.9cm}\includegraphics[angle=0,scale=0.28]{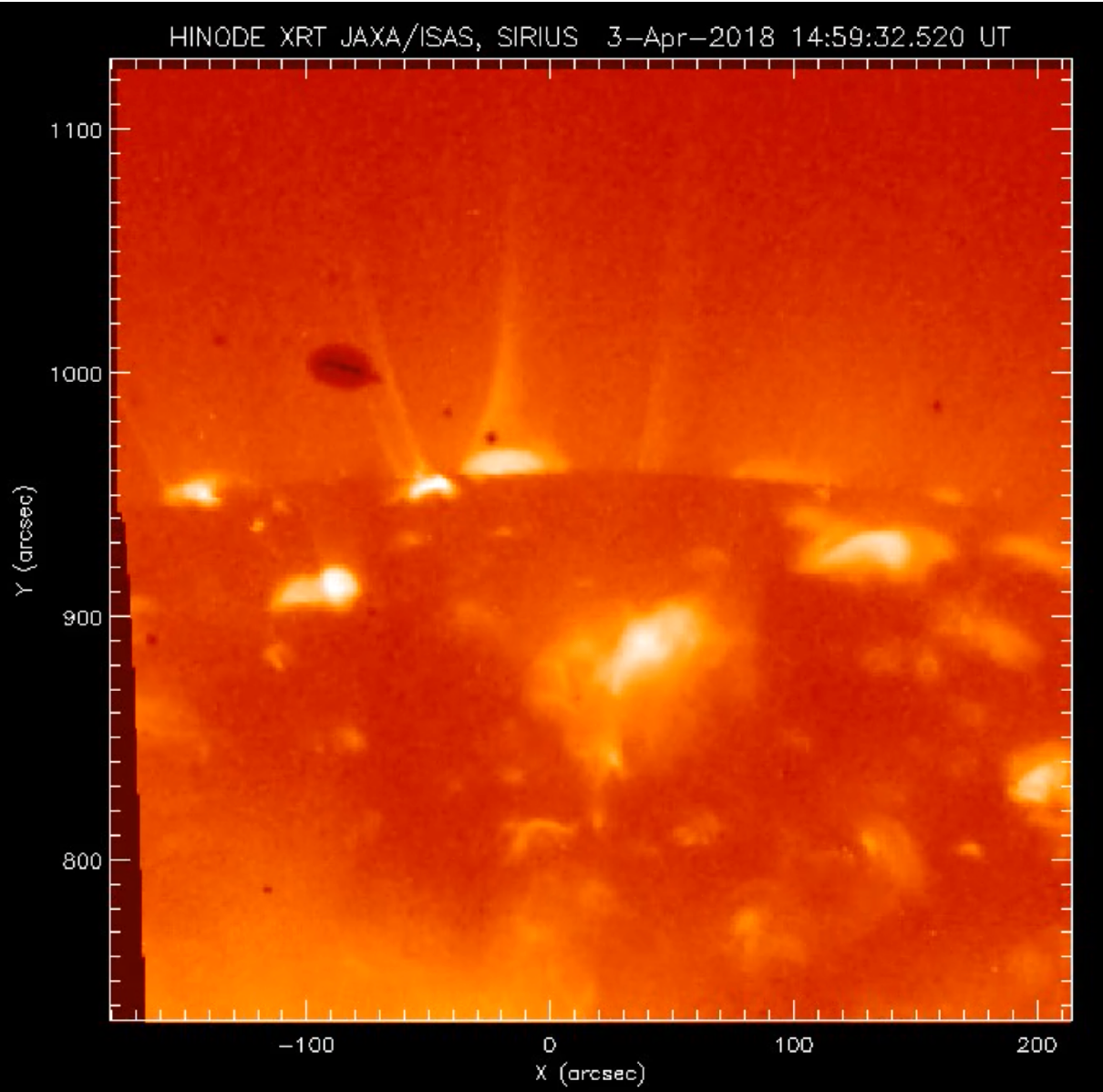}
\caption{The north pole region of the Sun in SXRs, from the X-Ray Telescope (XRT) on the Hinode spacecraft, with date and
time as in the top label.  Vertical outward ejections are coronal jets. (From Sterling et al.~2022, which gives more details.)} 
\end{figure} 

\vspace{0.5cm}

Jets apparently result from eruptions of small-scale filaments (minifilaments) (Fig. 2). This minifilament forms and erupts from a neutral 
line formed by a minority polarity patch surrounded by majority polarity. A jet bright point (JBP) often forms at the minifilament-eruption site (Sterling et~al.~2015). (Fig. 3.) Wyper et~al.~(2017) present numerical simulations of the basic process.
\begin{figure} 
\hspace*{0.2cm}\includegraphics[angle=0,scale=0.52]{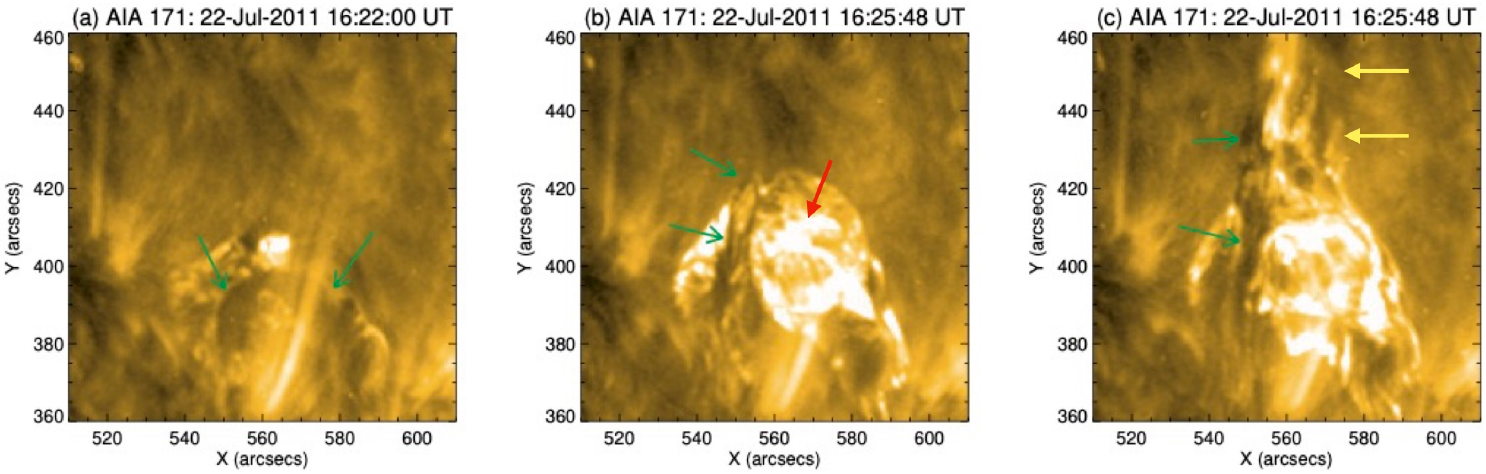}
\caption{Jet closeup from SDO/AIA 171Å channel, showing an erupting minifilament (green arrows), likely JBP (red), and the 
jet spire (yellow). (Sterling 2021; Shen et al.~2012.)} 
\end{figure}

\begin{figure} 
\hspace*{0.9cm}\includegraphics[angle=0,scale=0.44]{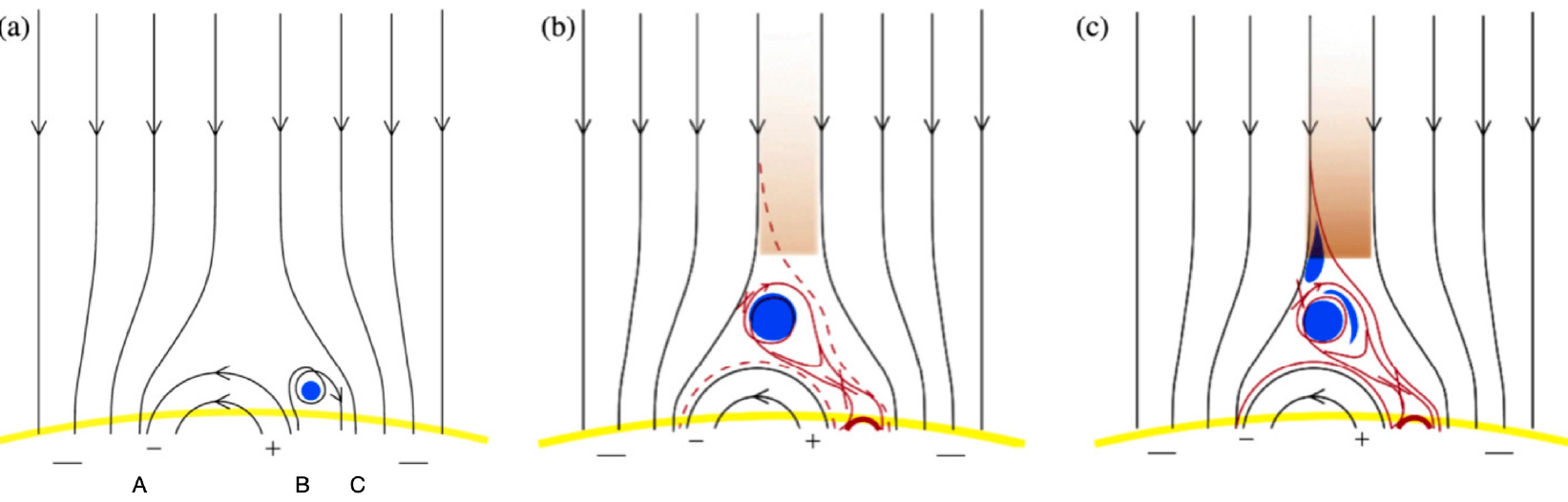}
\caption{Schematic picture of how jets work (Sterling et~al.~2015, 2018), showing magnetic field lines
before (black) and after (red) reconnection, erupting minifilament (blue), JBP (red semicircle), the jet spire (orange), and magnetic 
polarities (+, -). See Sterling et~al.~(2015) for details.} 
\end{figure}

The small-scale eruptions making jets are analogous to the eruptions that make CMEs (Fig. 4); the erupting minifilament and JBP in 
the jet case correspond to erupting large-scale filaments and a solar flare in the CME-eruption case.  We make a distinction between 
whether the erupting minifilament produces a "jet" or a "CME," based on whether the eruption results in the flux rope losing its 
flux-rope nature due to reconnection with the ambient coronal field.  In the case of a jet, one part of the erupting flux rope
completely reconnects with the external coronal field, so that eventually only one end of the erupting structure remains tied to the
photosphere.  In the case of a CME, the erupting flux rope has enough flux during its eruption so that part of the flux rope escapes
out into the heliosphere, with both ends of that flux rope remaining tied to the photosphere.  This point is also discussed in 
Sterling et al.~(2023).  It is, however, also possible for the eruptions that cause jets to trigger destabilization of a coronal loop to
create an erupting-loop CME, perhaps via the method suggested in Panesar et~al.~(2016a).

\begin{figure} 
\hspace*{0.0cm}\includegraphics[angle=0,scale=0.48]{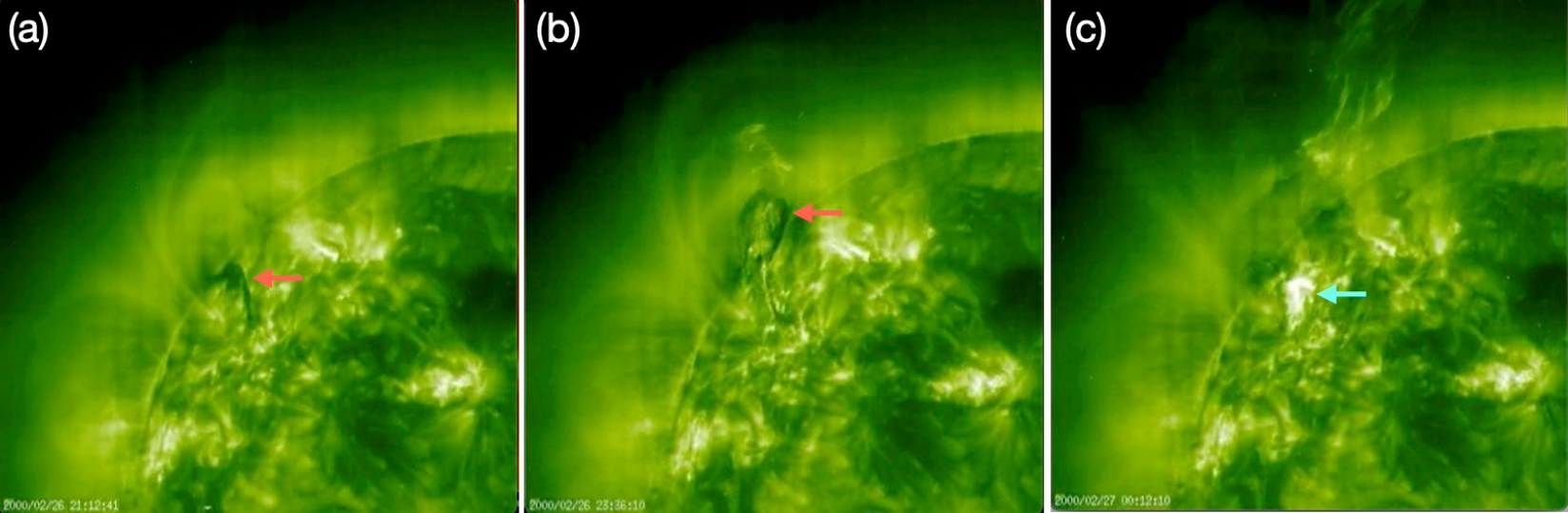}
\caption{Typical large-scale solar eruption. A solar filament (orange arrow) erupts outwards, in many cases eventually
producing a CME. A typical solar flare occurs in its wake (teal). Jets are apparetly small-scale versions of these large eruptions, with 
erupting minifilaments (JBP) analogous to erupting filaments (flares).  This eruption was analyzed in Sterling \& Moore (2004).} 
\end{figure} 

Other studies indicate that the jet-producing minifilament eruption is -- at least in many cases -- caused by magnetic flux cancelation in 
the photosphere (e.g., Panesar et al.~2016b, 2018; McGlasson et al.~2019; Muglach 2020).  Sometimes these jet-base 
regions can be small-scale ephemeral regions, as discussed in some detail in Moore et al.~(2022).  But not all jets originate
from such ephemeral regions that emerge as bipoles, and in fact in some cases 
the two opposite-polarity patches come from different, widely separated emergence locations, with one of the polarities 
migrating over and canceling with the other, resulting in the jet  (e.g., Adams et al. 2014).  In other cases, one polarity 
of the canceling neutral line can be the pole of an emerging ephemeral region, while the polarity with which it cancels 
is a pre-existing patch that developed independent of the emerging bipole (e.g., Panesar et al.~2018, 
Muglach~2021).  Schrijver~(2010) discussed eruptions from ephemeral regions that produce CMEs; although similar,
our jets differ from those eruptions in two ways: (1) jets are different from CMEs, as discussed 
above; and (2), not all of the jet-origin locations are EFRs, as discussed in this paragraph.

A figure in Shibata et al.~(1994) (their Fig. 2(b)) shows a proposed setup for jets with similarities to 
that in our Fig.~3, but there are key differences:  In ours the base magnetic lobe that moves outward
between our Fig.~3a and~3b erupts
explosively, with flaring reconnection below the erupting miniflament (resulting in the small bold red
semicircle loop between locations B and C, using the labels in Fig.~3a).
In the Shibata et al.~(1994) figure they write that lobe is ``emerging" or ``expanding," but there is no
mention that it is erupting and there is no internal flare reconnection depicted.  As a consequence, their 
case makes ``one bright loop and one jet," while ours makes two bright loops, represented by the two 
base loops in our Fig.~3c.   Our examination of observational data, especially of quiet Sun
and coronal hole jets (e.g., Sterling et al.~2015; Panesar et al.~2016b, 2018; McGlasson  et al.~2019), 
supports the view of our Fig.~3 and not that of the Shibata et al.~(1994) Fig. 2(b).  It is possible that the 
Shibata et al.~(1994) Fig.~2(b) scenario occurs in some different circumstances, and such evidence should
be presented if found.  (The Shibata et al. 1994 Fig.~2(c) case would represent a CME instead of a jet, in 
our view.)

\section{Coronal Jets and CME-producing Eruptions}

The jet-production mechanism may give insight into the onset of large-scale eruptions. One study (Sterling et~al.~2018) indicates 
that large-scale eruptions originating from magnetically isolated locations also result from flux cancelation, analogous to jets.  That 
study selected two active regions that could be followed on the solar disk from the time 
of emergence until the time when they produced an eruption.  Those two regions were both comparatively small ones that 
had a build-up time of about five days prior to eruption; larger regions typically evolve longer than two weeks (and so rotate 
onto or off the Earth-facing solar disk) before suddenly releasing the free energy that drives the resulting eruption.  Because 
the magnetic regions that produce jets are much smaller than even ``small" active regions, they progress much faster from the 
time of flux emergence to the time of jet production; Panesar et~al.~(2018) found this region-development time for jets to be 
from about 2 hours to two days prior to jet occurrence.  Similarly Moore et al.~(2022) found that ephemeral regions, some of 
which produced jets, lasted under two days.  

If, as we argue above, eruptions that produce jets are small-scale versions of larger eruptions, then we can possibly learn more 
about the lead up to and onset of eruptions from the eruptions that produce jets than from the larger eruptions.
This is because, in the case of jets, we can often use uninterrupted magnetogram coverage -- for about two days or less -- of the 
evolution of the magnetic field from its initial emergence until jet-producing-eruption onset. In contrast, it is harder to find large-scale
eruptions that build up quickly enough -- i.e.\ over less than or about two weeks -- for comparable studies of the pre-eruption 
magnetic flux evolution leading up to eruptions making typical flares/CMEs.  This opens the possibility of investigating general 
eruption ideas by statistically studying large numbers of jets, provided such studies can be carried out with adequate spatial 
resolution and cadence and at suitable wavelengths (Sterling et al.~2023). 

On the smaller size scale, the jet mechanism appears to produce features called ``jetlets" (Raouafi \& Stenborg~2014; 
Panesar et al.~2018, 2019; Kumar et~al.~2021; Sterling et al.~2023).  That same mechanism might also produce spicules, 
although this is more speculative (Sterling \& Moore 2016, Sterling et~al.~2020).

\section{Summary \& Discussion}

Solar coronal jets apparently result from small-scale eruptions analogous to CME-producing large-scale solar eruptions, with the minifilament eruption, JBP, and jet spire corresponding to a filament eruption, flare, and CME.

Jets are much more common than large-scale eruptions (several 100/day vs.\ $\sim$1/day), and they develop more quickly ($\sim$1-50 hrs vs.\ days/weeks (e.g., Yashiro et~al.~2004, Chen~2011, Panesar et~al.~2017, Sterling et~al.~2018).  

If jets and large eruptions work the same way, studies of the former can complement studies of the latter because of jets’ abundance and faster evolution (Sterling et al.~2023). Sterling~et~al.~(2018) provide examples of two small CME-producing active regions, where the eruptions apparently result
from processes analogous to those that cause jets.  By further clarifying and confirming the onset mechanism for the comparatively frequent jets,
we potentially can gain insight into the onset mechanism for the larger-scale eruptions also.   

Moreover, CMEs on other stars likely occur via the same mechanism as CMEs from the Sun (e.g., Veronig et~al.~2021).  Therefore, exploration of
among the smallest eruptive features on the Sun (the small-scale minifilament/flux-rope eruptions that make coronal jets) potentially can 
provide clarification for far-away stellar eruptions also.

\vspace{0.5 cm}
 The authors received funding from the Heliophysics Division of NASA's Science 
Mission Directorate through the Heliophysics Supporting Research (HSR, grant No.~20-HSR20\_2-0124) Program, 
and the Heliophysics Guest Investigators program.  ACS and RLM were also supported 
through the Heliophysics System Observatory Connect (HSOC, grant No.~80NSSC20K1285) 
Program, and A.C.S. received support from the NASA/MSFC Hinode satellite Project Office.  N.K.P. received 
additional support through a NASA SDO/AIA grant.   
We acknowledge the use of AIA data. AIA is an instrument onboard SDO, a mission of
NASA's Living With a Star program.

\vspace{1 cm}

\noindent{\bf Q \& A}

\vspace{0.1 cm}

\noindent {\bf Philippe L. Lamy}: In our 2019 review paper (Lamy, P., et al.~2019, Space Sci.~Rev., 215, 39) we argue that CMEs at large
arise from closed-field coronal regions at both large and small size scales.  Your presentation appears consistent with our argument.

\vspace{0.3 cm}

\noindent {\bf A. C. Sterling}:  I largely agree with your assessment, but with some caveats on the creation of the smallest CMEs.  
We have argued that on the size scale of jets and smaller there is a physical difference between jets and CMEs.  In our jet picture 
(Fig.~3), the erupting flux rope carrying the erupting minifilament is totally consumed through reconnection with the open field 
(Figs.~3b and~3c), with the twist of the initially twisted minifilament/flux-rope unravelling onto the open field, and propagating 
out along the field as an Alfvenic pulse (Shibata, K. and Uchida, Y. 1986, Solar Phy.\ 178, 379; Sterling, A. \& Moore, R. 2020, 
ApJ, 896, L18).  Thus, in that jet case, a jet spire forms in the corona, and a ``white-light jet" or ``narrow CME" (e.g., Wang, Y.-M., 
et al.~1998, ApJ, 508, 899; Moore et al.~2015, ApJ, 806, 11) can form in the outer corona rather
than a {\it bona fide} CME\@.  Such {\it bona fide} CMEs, on the other hand, are generally thought of as having a flux-rope core.  In order 
to make such a small-scale CME from a jet-like magnetic setup of Fig.~3, an erupting small-scale flux rope would have to be
robust enough to survive as a flux rope the interchange reconnection of Figs.~3b and 3c.  So in 
summary: below some (rather ill-defined) cutoff size the erupting minifilament/flux-rope will make a coronal jet and a narrow CME,
and above that cutoff size a portion of the erupting minifilament/flux-rope will escape as a flux rope -- perhaps severely reduced in 
flux content -- and potentially form the core of a very small CME\@.  There would be some sort of mixture of the features for erupting
miniflament/flux-ropes of size close to that cutoff size.  We discuss some of these points further in Sterling et al.~(2023).  Additionally,
some CMEs might be triggered via the mechanism discussed in Panesar~et~al.~(2026a), whereby the twist of a growing jet's field gets
transferred to the base of a coronal loop, and subsequently destabilizes that loop causing it to erupt.

\end{document}